\def\editmode{0}

\def\hyperlinks{0}  

\def\spsformat{0}

\def\singlenarrowcol{0} 

\def\bibfilenames{bibman_refs}

\def\journal{0}

\def\intermediatesteps{1}

\if\spsformat1

\documentclass{article}
\usepackage{spconf}

\else

\if\singlenarrowcol0
\documentclass[conference]{IEEEtran}

\IEEEoverridecommandlockouts
\else

\documentclass[onecolumn]{IEEEtran}
    \usepackage[pass]{geometry}
    \newgeometry{textwidth=252pt, textheight=672pt}

\fi
\fi

\usepackage{include_v8}
\usepackage{notation}

\newcommand{\nextversion}[1]{}

\begin{document}

\if\spsformat1
    \title{Title}
    \name{Author(s) Name(s)\thanks{Thanks to XYZ agency for funding.}}
    \address{Author Affiliation(s)}
\else
    \title{Learning the Channel Gain \\from Anywhere to Anywhere via \\Cross-environment Transformer Estimators
        \thanks{This work has been funded by the IKTPLUSS grant 311994 of the Research
            Council of Norway.}}

    \author{
        Prasenjit Dhara and Daniel Romero
        \\
        Dept. of Information and Communication Technology,
        University of Agder, Grimstad, Norway.\\
        Email:\{prasenjit.dhara,daniel.romero\}@uia.no.
    }

\fi

\maketitle

\begin{abstract} 
    Channel-gain maps provide the channel gain between any two locations in a geographical region. They find numerous applications, from resource allocation and interference control to path planning for autonomous vehicles. Channel-gain map estimation (CGME) is considerably more challenging than conventional radio map estimation (RME) because channel-gain maps are functions over a 6-dimensional input space. This calls for specialized methods, which currently rely  on the (inaccurate) radio tomographic model or require a prohibitively large number of measurements since they do not exploit any spatial structure. This paper overcomes this issue by  leveraging spatial patterns that channel-gain maps exhibit across environments, as dictated by the laws of physics and typical environmental characteristics (e.g. building materials and layouts). Adopting a metalearning perspective, a transformer-based estimator is proposed to implicitly  learn this common structure from measurements collected in multiple environments. This enables CGME in new environments from  significantly fewer  measurements (five times less in our experiments). To maximize learning efficiency, the transformer is composed with a  feature map that enforces the invariances of CGME, such as those following from reciprocity. Numerical experiments corroborate the merits of the proposed estimator relative to existing methods.
\end{abstract}

\newcommand\blfootnote[1]{%
    \begingroup
    \renewcommand\thefootnote{}\footnote{#1}%
    \addtocounter{footnote}{-1}%
    \endgroup
}

\if\spsformat0
    \begin{IEEEkeywords}
        Radio map estimation, channel-gain maps, transformers, cross-environment learning.
    \end{IEEEkeywords}
\fi

\section{Introduction}
\label{sec:intro}
\begin{bullets}%
    \blt[overview]
    \begin{bullets}%
        \blt[radio maps]
        \begin{bullets}%
            \blt[overview]Radio maps characterize the propagation or occupancy state of a radio communication channel in a geographical region~\cite{romero2022cartography,romero2024theoretical}.
            \blt[channel gain radio maps]A particularly important class of radio maps is the so-called \emph{channel-gain maps}, which provide the channel gain between any two locations in the region of interest.
            \blt[applications]Clearly, such information is of great value, as it can be used  for resource allocation, interference control, routing in ad-hoc networks, scheduling in device-to-device communications, aerial base station deployment, and path planning for swarms of autonomous vehicles, among many other applications; see e.g.~\cite{romero2022aerial,romero2016channelgain}. An illustrative example is that of a wireless network with $N$ terminals: measuring the channel gain between each pair of terminals requires $N(N-1)/2$ measurements. In turn, using a channel-gain map, estimates of those values can be obtained directly from the locations of the terminals.
        \end{bullets}%

        \blt[RME]
        \begin{bullets}%
            \blt[Overview]Radio map estimation (RME) deals with the construction of radio maps from  measurements and their locations~\cite{romero2022cartography}.
            \blt[CGME]This paper focuses on the  construction of channel-gain maps, which is referred to as \emph{channel-gain map estimation} (CGME).
        \end{bullets}%
    \end{bullets}%

    \blt[Related work]
    \begin{bullets}
        \blt[Non-CGME RME]The literature of RME is sizable and rapidly growing.
        \begin{bullets}
            \blt[before DL]
            \begin{bullets}
                \blt[kernel]Estimators were developed based on
                \blt[kriging] Kriging~\cite{alayafeki2008cartography,\nextversion{agarwal2018spectrum,}shrestha2022surveying},
                \blt[kbl]kernel-based learning~\cite{romero2017spectrummaps,teganya2019locationfree},
                \blt[Matrix completion]matrix completion~\cite{schaufele2019tensor\nextversion{,khalfi2018airmap}},
                \blt[sparsity]sparsity-based inference~\cite{bazerque2010sparsity\nextversion{,bazerque2011splines,jayawickrama2013compressive}},                %
                \blt[dictionary learning] dictionary learning~\cite{kim2013dictionary}, graphical models~\cite{ha2024location},
            \end{bullets}
            \blt[deep learning]
            \begin{bullets}%
                \blt[DNNs and CNNs]convolutional neural networks (CNNs)~\cite{krijestorac2020deeplearning,levie2019radiounet,teganya2020rme,shrestha2022surveying}, and,
                \blt[Transformers]more recently, transformers~\cite{viet2025spatial}; see also~\cite{pandey2021limited,wang2023bert}.
            \end{bullets}
        \end{bullets}
        \blt[CGME]%
        \begin{bullets}%
            \blt[why CGME is different]Unfortunately, these methods cannot be generally applied to estimate channel-gain maps, as these maps are functions of two locations rather than one. 
            \begin{journalonly}                        
            Some of the above methods, such as those based on CNNs, are  inapplicable by construction. Others, such as those based on kernel methods, could in principle be applied, but a reasonable performance cannot be expected due to their highly limited learning capabilities.
            \end{journalonly}
            \blt[specialized methods]Instead, specialized methods for CGME are required.

            \begin{bullets}%
                \blt[tomographic]By far, the most popular approach to CGME relies on the so-called \emph{radio tomographic model}, which expresses the channel gain as a path-loss term minus the line integral of a \emph{spatial loss field} that quantifies absorption at each point~\cite{patwari2008nesh,agrawal2009correlated,
                    wilson2009regularization,kanso2009compressed,dallanese2011kriging,romero2016channelgain,lee2016lowrank,lee2018adaptive,romero2022aerial}. Unfortunately, this approach suffers from large modeling error, as propagation in typical communication bands is not dominated by absorption.
                    \begin{journalonly}  
                    but by reflection, diffraction, and scattering, which are not captured by the tomographic model.
                    \end{journalonly}
                \blt[non-tomographic]Pure  data-driven approaches sidestep this issue. They comprise methods based on\footnote{Note the existence of works on related but different problems, such as \emph{pathloss prediction} (see e.g. \cite{lee2024pathloss}) where channel gain is approximated given environmental features rather than measurements. Also, some recent papers  use the term ``channel-gain map" with a different meaning~\cite{sun2025scatterer}.}
                \begin{bullets}%
                    \blt[spatiotemporal filtering]spatiotemporal filtering~\cite{kim2011kriged},
                    \blt[kernel]adapted kernel-based learning~\cite{lopezramos2020moe},
                    \blt[deep learning]and deep neural networks (DNNs) \cite{imai2019radiopredictioncnn,iwasaki2020transferbasedpower}.
                \end{bullets}
            \end{bullets}%
        \end{bullets}
        \blt[existing gap]Unfortunately, since the target function is defined on a 6-dimensional space and these approaches do not leverage any prior information, they require a prohibitive number of measurements. 
    \end{bullets}
    
    \blt[contribution]
    \begin{bullets}%
        \blt[This paper]To overcome this issue, the key observation in this paper is that a channel-gain map is not just \emph{any}  function: it exhibits a certain spatial structure dictated by the laws of physics (namely Maxwell's equations) and manifests common patterns (e.g. the channel gain generally decreases with distance and is often reduced by obstacles).
        \blt[data-driven]The main contribution of this paper is a novel approach where this common structure is \emph{implicitly  learned} from measurements collected in multiple environments. By doing so, estimating channel-gain maps in new environments can be accomplished with significantly fewer measurements than what existing methods require (5 times less in the setup of Sec.~\ref{sec:experiments}), since an important part of the information has already been learned offline. 

        \blt[transformer]The proposed \emph{cross-environment transformer-based estimator} (CRETE) materializes this approach via a DNN architecture based on the composition of a transformer and a feature map that enforces the  invariances of the CGME problem. These invariances follow from reciprocity and the geometrical symmetries of Maxwell's equations. 
        \blt[novelty]To the best of our knowledge, this is the first work to (i) consider cross-environment learning for CGME, and to (ii) utilize a transformer-based estimator for CGME.
    \end{bullets}%

    \blt[paper structure]Sec.~\ref{sec:problem} introduces the model and formalizes the problem. Sec.~\ref{sec:transformers} reviews the core concepts behind transformers. Sec.~\ref{sec:proposed} 
    develops cross-environment estimation and proposes the transformer-based estimator. Finally, Sec.~\ref{sec:experiments} presents numerical results and Sec.~\ref{sec:conclusions} concludes the paper. The code and data required to reproduce the experiments can be found at \url{www.radiomaps.org}.

\end{bullets}%


\section{Model and Problem Formulation}
\label{sec:problem}

\begin{bullets}%
    \blt[model]
    \begin{bullets}%
        \blt[region] Let $\Reg\subset\rfield^{3}$ denote the set of all  points within the region of interest.
        \blt[channel model]
        \begin{bullets}%
            \blt[propagation] The signal $\TraSig(t)$ transmitted by a terminal at $\TraLoc\in\Reg$ propagates through the environment to a terminal at $\RecLoc\in\Reg$.
            \blt[Rx. signal]The received signal  can be expressed  as $\RecSig(t)=\Cha(t)\ast \TraSig(t) + \Noi(t)$, where $\Cha(t)$ is the channel impulse response and $\Noi(t)$ is noise. In the frequency domain, this reads as $\FouRecSig(f)=\FouCha(f) \FouTraSig(f) + \FouNoi(f)$.
        \end{bullets}%
        \blt[map]
        \begin{bullets}%
            \blt[freq]The
            \emph{channel gain} at a given frequency $\FreFix$ is defined as $10 \log_{10}|\FouCha(\FreFix)|^2$.
            \blt[isotropic]If both terminals have isotropic antennas and the environment does not change over time, this gain depends only on $\TraLoc$ and $\RecLoc$. As a result, it can be formalized as a function $\Map:\Reg \times \Reg \rightarrow \rfield$ referred to as a \emph{channel-gain map}.
            \blt[reciprocity]Note that, due to reciprocity, $\Map(\TraLoc,\RecLoc)=\Map(\RecLoc,\TraLoc),~\forall\TraLoc,\RecLoc\in\Reg$.
            \blt[terminal-independence]
            \blt[Propagation factors]
        \end{bullets}%

        \blt[measurement model]To estimate $\Map$, the channel gain is measured between  $\MeaNum$
        location pairs $\left\{(\TraLoc_{\MeaInd},\RecLoc_{\MeaInd}) \right\}_{\MeaInd=1}^{\MeaNum}
            \subset \Reg^2$. This can be accomplished e.g. by static terminals spread across $\Reg$ or by a small number of terminals that move to different locations. In any case, the $\MeaInd$-th measurement can be written as
        $
            \Mea_{\MeaInd} = \Map(\TraLoc_{\MeaInd},\RecLoc_{\MeaInd})  + \MeaNoi_{\MeaInd}$,
        where $\MeaNoi_{\MeaInd}$ denotes measurement error.
    \end{bullets}%

    \blt[Problem Formulations]
    \begin{bullets}
        \blt[given] Given the measurements $\MeaSet\define\left\{(\TraLoc_{\MeaInd},\RecLoc_{\MeaInd}, \Mea_{\MeaInd}) \right\}_{\MeaInd=1}^{\MeaNum}
        $
        \blt[requested]the CGME problem is to estimate the function $\Map$ or, equivalently, the values $\Map(\TraLoc,\RecLoc)~\forall \TraLoc,\RecLoc\in\Reg$.
    \end{bullets}
\end{bullets}%

\section{Background on Transformers}
\label{sec:transformers}

\begin{bullets}%
    \blt[overview]Transformers are a type of DNN architecture based on \emph{inner-product attention}~\cite{vaswani2017attention}.
    \blt[architecture]Internally, transformers are the result of composing \emph{attention layers}, multilayer perceptrons, and normalization layers with residual connections.
    \blt[function]Importantly, a  transformer can be seen as a function $\TraFun$ that takes a set of $\QueVecNum$ vectors as input and produces a set of $\QueVecNum$ vectors at the output. Formally,
    $\TraFun: \rfield^{\EmbDim \times \QueVecNum}\rightarrow \rfield^{\EmbDim \times \QueVecNum}$, and the same network can be applied regardless of $\QueVecNum$.
    \blt[permutation equivariance]Function $\TraFun$  satisfies \emph{permutation equivariance}, meaning that permuting  the columns of $\QueMat$ permutes the columns of $\TraFun(\QueMat)$ in the same~way. This is useful here because measurement sets are unordered and their cardinality $\MeaNum$ varies across deployments.
\end{bullets}%

\section{Cross-environment CGME}
\label{sec:proposed}

\begin{bullets}%
    \blt[Overview]This section develops the proposed estimation approach. The starting point is to theoretically investigate why existing methods cannot effectively produce accurate channel-gain map estimates. This motivates a \emph{cross-environment estimation approach} (Sec.~\ref{sec:crossenvlearning}), which is subsequently generalized into an \emph{implicit learning formulation} (Sec.~\ref{sec:implicitlearning}). A transformer-based estimator materializes this framework in Sec.~\ref{sec:transformerimpl}.

    \blt[when CGME is needed]The first step when tackling the CGME problem is to determine whether a CGME-specific approach is actually needed. If one terminal location is fixed, conventional RME tools (which estimate functions of a single spatial variable) may suffice~\cite{romero2022cartography}; genuine CGME estimators are needed when generalization  \emph{across both locations} is required. 

    \blt[Per-environment CGME]With this in mind, a natural approach to CGME is \emph{parametric estimation}, which seeks a map estimate in a family of functions $\{\Map_\EnvPar,~\EnvPar\in\EnvParSet\}$. Here, $\EnvPar$ is a vector of parameters (e.g., the weights and biases of a neural network) and $\EnvParSet$ the set of allowable values of $\EnvPar$.
    \begin{bullets}%
        \blt[criterion]Estimation will generally be accomplished by solving
        \begin{align}
            \label{eq:singleenv}
            \minimize_{\EnvPar\in\EnvParSet}\frac{1}{\MeaNum}\sum_{\MeaInd=1}^{\MeaNum}(\Mea_{\MeaInd}-\Map_\EnvPar(\TraLoc_\MeaInd,\RecLoc_\MeaInd))^2.
        \end{align}
        A square loss was adopted in this and subsequent expressions for illustration purposes, but any other loss can be considered.   

        \blt[examples]
        \begin{bullets}%
            \blt[Tomographic]Prominent estimators of this kind are based on the radio tomographic model, where $\EnvPar$ represents the spatial loss field and $\Map_\EnvPar(\TraLoc,\RecLoc)$ is its line integral between $\TraLoc$ and $\RecLoc$ plus path loss; cf. Sec.~\ref{sec:intro}. This model captures absorption by matter (e.g. buildings), but cannot accommodate reflection, diffraction, and other propagation phenomena.
            \blt[Data-driven]Indeed, model-based approaches are unlikely to yield high-quality estimates (especially when the amount of data is low) due to the high complexity of electromagnetic propagation~\cite{romero2024theoretical}.   Instead, data-driven approaches are preferable since they do not introduce inaccurate assumptions on $\Map$~\cite{kim2011kriged,lopezramos2020moe,imai2019radiopredictioncnn,iwasaki2020transferbasedpower}.

        \end{bullets}%
    \end{bullets}%

\end{bullets}%

\subsection{Cross-environment Learning}
\label{sec:crossenvlearning}

\begin{bullets}%
    \blt[motivation]
    \begin{bullets}%
        \blt[single environment ok if much data]The approach discussed so far is suitable provided that $\MeaSet$ is  sufficiently informative about $\Map$. 
        \blt[grid]However, this is unlikely as $\Map$ is a function defined on a 6-dimensional space. For illustration purposes, note that a grid of (only) 10 points per dimension has $10^6$ points in total. This is the so-called \emph{curse of dimensionality}~\cite{scholkopf2001}. 
        \blt[limited data in practice]In contrast, practical setups may require the estimation of  $\Map$ from just a few hundred measurements. For example, in an environment with 40 static terminals, at most $40(40-1)/2=780$ measurements will be available.
        \blt[too much variance]Thus, practical CGME is hopeless unless one exploits further structure. 
        \blt[side information]For this reason,  \cite{imai2019radiopredictioncnn,iwasaki2020transferbasedpower} capitalize on side information such as building or foliage maps. However, in the standard CGME setup, such information is not available.
    \end{bullets}%

    \blt[Learning across environments]
    The key idea in this paper is to  sidestep this difficulty by considering \emph{multiple instances} of the CGME problem.
    \begin{bullets}%
        \blt[datasets]Specifically, suppose that one is given $\EnvNum$ measurement sets $\MeaSet_{1},\ldots, \MeaSet_{\EnvNum}$, each one from a different environment (geographical region).
        \blt[common+specific]Clearly, each of these sets contains information that is 
        \emph{common} to all environments and information that is \emph{specific} to its environment.
        \begin{bullets}
            \blt[common]Intuitively, the  common information pertains to the way radio waves propagate across space and interact with matter, whereas
            \blt[specific]the specific information in $\MeaSet_\EnvInd$  encodes the special characteristics of the $\EnvInd$-th environment. This includes the positions, shapes, and materials of the obstacles, such as buildings, vegetation, and terrain features.
        \end{bullets}%
        \blt[intuition] Plain data-driven approaches based on \eqref{eq:singleenv} fail if the amount of information in $\MeaSet$ is insufficient to learn \emph{both} the common and specific information. However, if one can learn the common information offline from measurements in other environments, estimating $\Map$ in a new environment just requires learning the specific information, which can be accomplished with a much smaller~$\MeaSet$. 

    \end{bullets}%

    \blt[Overview]Having motivated \emph{cross-environment learning for CGME}, the rest of this section describes how this idea can be brought into practice by building upon notions from \emph{few-shot learning} and \emph{metalearning}~\cite{finn2017maml}.
    \blt[Common+specific parameters]To this end, it is instructive to expand the earlier parametric representation into the functions $\{\Map_{\ComPar,\EnvPar},~\ComPar\in\ComParSet, \EnvPar\in\EnvParSet\}$, where $\ComPar$ encodes the common information and $\EnvPar$ the environment-specific information.  The datasets $\MeaSet_{1},\ldots, \MeaSet_{\EnvNum}$ will be used to obtain a single $\ComPar$ as well as $\EnvNum$ different $\EnvPar_1,\ldots,\EnvPar_{\EnvNum}$, one per environment.

    \blt[estimation approach]
    \begin{bullets}%
        \blt[Finding specific parameters given common \ra minimize]To see how this can be done, suppose that this operation has been accomplished (so $\ComPar, \EnvPar_1,\ldots,\EnvPar_{\EnvNum}$ are known) and 
        consider the problem of finding the environment-specific parameters $\EnvPar$ for a new environment given  $\MeaSet\define\left\{(\TraLoc_{\MeaInd},\RecLoc_{\MeaInd}, \Mea_{\MeaInd}) \right\}_{\MeaInd=1}^{\MeaNum}
        $.
        \begin{bullets}%
            \blt[split] The standard machine learning approach would be  to split $\MeaSet$ into a training $\TraMeaSet$ and a validation set $\ValMeaSet$. Formally, let $\TraMeaSet\define\left\{(\TraLoc_{\MeaInd},\RecLoc_{\MeaInd}, \Mea_{\MeaInd}),~\MeaInd\in \TraMeaIndSet \right\}$ and $\ValMeaSet\define\left\{(\TraLoc_{\MeaInd},\RecLoc_{\MeaInd}, \Mea_{\MeaInd}),~\MeaInd\in \ValMeaIndSet \right\}$, where $\TraMeaIndSet$ and $\ValMeaIndSet$ are disjoint index sets.
            \blt[objective]Given this split, one can  aim to find
            \begin{salign}
                \label{eq:envparopt}
                \EnvParOpt &= \argmin_{\EnvPar\in\EnvParSet} \frac{1}{|\TraMeaIndSet|} \sum_{\MeaInd\in \TraMeaIndSet}\Big(\Mea_{\MeaInd}-\Map_{\ComPar,\EnvPar}(\TraLoc_\MeaInd,\RecLoc_\MeaInd)\Big)^2 \\&\define \EnvPar(\TraMeaSet;\ComPar),
            \end{salign}
            where $|\MeaIndSet|$ denotes the cardinality of set $\MeaIndSet$.
        \end{bullets}%
        \blt[Quantifying performance]The quality of  $\EnvParOpt$ can be quantified by the validation loss
        \begin{salign}
            \label{eq:valenvparopt}
            \LosFun(\ValMeaSet, \EnvParOpt) &= \frac{1}{|\ValMeaIndSet|} \sum_{\MeaInd\in \ValMeaIndSet}\Big(\Mea_{\MeaInd}-\Map_{\ComPar,\EnvParOpt}(\TraLoc_\MeaInd,\RecLoc_\MeaInd)\Big)^2\\&=\LosFun(\ValMeaSet, \EnvPar(\TraMeaSet;\ComPar)).
        \end{salign}
    \end{bullets}%

    \blt[Finding common parameters]Thus far, the approach to find $\EnvPar$ given $\ComPar$ was described. It remains to find the common parameters $\ComPar$.
    \begin{bullets}%
        \blt[criterion]To this end, note that the quality of a given $\ComPar$ can be quantified by the average across environments of the validation loss resulting from $\EnvPar(\TraMeaSet_{\EnvInd};\ComPar)$. Thus, a sensible criterion to find $\ComPar$ is
        \begin{align}
            \label{eq:comparopt}
            \ComParOpt = \argmin_{\ComPar\in\ComParSet} \frac{1}{\EnvNum} \sum_{\EnvInd=1}^{\EnvNum} \LosFun(\ValMeaSet_{\EnvInd},
            \EnvPar(\TraMeaSet_{\EnvInd};\ComPar)).
        \end{align}
        \blt[challenging optimization]
        \begin{bullets}%
            \blt[closed]Solving  \eqref{eq:comparopt} is not generally feasible since $\EnvPar(\TraMeaSet_{\EnvInd};\ComPar)$ is the minimizer of \eqref{eq:envparopt}, which typically has no closed form.
            \blt[simultaneous]Jointly minimizing over  $\ComPar$ and $\EnvPar_1,\ldots,\EnvPar_{\EnvNum}$ would violate the information flow since the resulting  $\EnvParOpt_1,\ldots,\EnvParOpt_{\EnvNum}$ would depend on $\ValMeaSet_1,\ldots,\ValMeaSet_{\EnvNum}$.
        \end{bullets}%

        \blt[few steps \ra MALM/Cavia]In the metalearning literature, a common approach to bypass this difficulty is to replace the exact optimization in \eqref{eq:envparopt} with a few steps of an optimization algorithm such as gradient descent~\cite{finn2017maml,zintgraf2019cavia}.
        \blt[Encoder \ra CNP/ANP]Another approach is to step back and realize that $\EnvPar(\TraMeaSet;\ComPar)$ is ultimately a function of the measurements in $\TraMeaSet$  and $\ComPar$. This function can be approximated by a DNN and substituted into \eqref{eq:comparopt} to find $\ComPar$, which now includes also the parameters of this DNN; see e.g.~\cite{garnelo2018cnp}.
    \end{bullets}%

    \blt[Implicit \ra exploit variable inputs/permutation equivariance of transformers \ra transformer]Although these approaches yield functional CGME estimators, the inaccuracies introduced by the aforementioned approximations are oftentimes greater than the inaccuracy introduced by adopting the tomographic model. To study this effect, we developed and tested several implicit estimators, some of them described in Sec.~\ref{sec:experiments}. The insufficient performance improvement relative to prior art that was observed motivates the approach in the next section, which leverages the great flexibility of transformers to compensate for this effect.

\end{bullets}%

\subsection{Implicit Cross-environment Learning}
\label{sec:implicitlearning}

\begin{bullets}%
    \blt[Explicit learning]
    \begin{bullets}%
        \blt[Explicit estimator form]Observe that the approach in the previous section yields map estimates of the form
        \begin{align}
            \label{eq:explicitest}
            \MapEst(\TraLoc,\RecLoc) = \Map_{\ComPar,\EnvPar(\TraMeaSet;\ComPar)}(\TraLoc,\RecLoc).
        \end{align}
        \blt[interpretation]The right-hand side depends on $\TraMeaSet$ only through $\EnvPar(\TraMeaSet;\ComPar)$, which can therefore be seen as a vector summarizing the relevant information in $\TraMeaSet$. Thus, $\EnvPar(\TraMeaSet;\ComPar)$ captures the impact of the environment on propagation in the same way as a spatial loss field captures the absorption at each grid point.
        \blt[Naming]For this reason,  estimators of the form \eqref{eq:explicitest} will be referred to as \emph{explicit} cross-environment learning estimators.

    \end{bullets}%

    \blt[Implicit learning]
    \begin{bullets}%
        \blt[Overview]In contrast, the rest of this section explores an alternative perspective, where the environment-specific information is not explicitly encoded.
        \blt[Estimator form]To this end, step back to   notice that the right-hand side of
        \eqref{eq:explicitest} is  a function of $\TraLoc,\RecLoc,\TraMeaSet$, and $\ComPar$, so a more general form would be
        \begin{align}
            \label{eq:implicitest}
            \MapEst(\TraLoc,\RecLoc) = \Map_{\ComPar}(\TraLoc,\RecLoc; \TraMeaSet).
        \end{align}
        \blt[generalization]Note that all estimators of the form in \eqref{eq:explicitest} are also of the form in \eqref{eq:implicitest}, but not vice versa. In particular, the form in \eqref{eq:implicitest} allows for a more general dependence on $\TraMeaSet$ that is not necessarily captured by a single vector $\EnvPar(\TraMeaSet;\ComPar)$.
        \blt[term]Estimators that can be written as in \eqref{eq:implicitest} but not as in \eqref{eq:explicitest} will be referred to as \emph{implicit} cross-environment learning estimators.

        \blt[training]The upside of implicit estimators is that they admit a more natural training procedure, since  $\ComPar$ can be obtained from
        \begin{align}
            \label{eq:implicitcomparopt}
            \minimize_{\ComPar\in\ComParSet} \frac{1}{\EnvNum} \sum_{\EnvInd=1}^{\EnvNum}\frac{1}{|\ValMeaIndSet_{\EnvInd}|} \sum_{\MeaInd\in \ValMeaIndSet_\EnvInd}\Big(\Mea_{\EnvInd,\MeaInd}-\Map_{\ComPar}(\TraLoc_{\EnvInd,\MeaInd},\RecLoc_{\EnvInd,\MeaInd}; \TraMeaSet_{\EnvInd})\Big)^2.
        \end{align}
        Here, the subscript notation has been extended to also indicate  the environment index.
        \blt[Resampling environments]Note that the loss in \eqref{eq:implicitcomparopt} has been presented with $\EnvNum$ outer summands for simplicity. In practice, each environment $\MeaSet_{\EnvInd}$ is randomly resampled into $\TraMeaSet_{\EnvInd}$ and $\ValMeaSet_{\EnvInd}$ at each training step where this environment is selected.
       
    \end{bullets}%
\end{bullets}%
\subsection{Invariance Preserving Transformer-based Estimator}
\label{sec:transformerimpl}
\begin{bullets}%
    \blt[Motivation transformers]It remains to provide a means to implement estimators of the form $\Map_{\ComPar}(\TraLoc,\RecLoc; \MeaSet)$. As seen next, transformers are highly suitable for this task because  $\MeaSet$ contains a variable number $\MeaNum$ of measurements; cf.  Sec.~\ref{sec:transformers}. 
    \blt[estimator form]Since a transformer takes matrix inputs and returns matrix outputs, proper adaptation functions are required. Specifically, the targeted estimator will be constructed as  the composition $\Map_{\ComPar}(\TraLoc,\RecLoc; \MeaSet) = \OutFun(\TraFun_{\ComPar}(\InpFun_{\ComPar}(\TraLoc,\RecLoc; \MeaSet)))$, where $\InpFun_{\ComPar}$ maps the input to a matrix, $\TraFun$ is a transformer, and $\OutFun$ maps the output of the transformer to a scalar. 
    

    \blt[Design of $\InpFun_{\ComPar}$]
    \begin{bullets}%
        \blt[Not every choice OK]To achieve good estimation performance, it is convenient to attune $\InpFun_{\ComPar}$ to 
        \blt[structure \ra Invariances]the \emph{invariances} and \emph{equivariances}~\cite{lehmann} of the CGME problem. For example, changing the order of the measurements in $\MeaSet$ should not affect the map estimate. A poor choice of $\InpFun_{\ComPar}$ forces the transformer to learn these invariances from data, which is inefficient, dramatically increases the data needs, and limits the ability of the estimator to learn complex patterns.
        \blt[approach here]A suitable design is pursued here by forming $\InpFun_{\ComPar}$ as a composition of transformations $\InpFun_1$, $\InpFun_2$, \ldots,  each one rendering $\InpFun_{\ComPar}$ (and therefore $\Map_{\ComPar}$) invariant to one of the invariances of the problem. Each of these transformations is briefly discussed next (details can be found in the code):  
        \blt[Exploitation of the invariances]
        \begin{itemize}
            \item$\InpFun_1$ (\textbf{Symmetry with respect to the evaluation locations}): The first transformation satisfies $\InpFun_1(\TraLoc,\RecLoc; \MeaSet) = \InpFun_1(\RecLoc,\TraLoc; \MeaSet)$. To this end, it returns  $(\TraLoc,\RecLoc, \MeaSet)$ if $\TraLoc$ is on average closer to the points in $\MeaSet$ than $\RecLoc$, and
            $(\RecLoc,\TraLoc, \MeaSet)$ otherwise.


            \item$\InpFun_2$ (\textbf{Invariance to horizontal translation}): $\InpFun_2$ shifts the origin to $\TraLoc$ by subtracting  the coordinates of $\TraLoc$ from  all points in $(\TraLoc,\RecLoc, \MeaSet)$. Since the original  height of $\TraLoc$ is informative about the distance to the ground, it is retained as a separate feature. 

            \item$\InpFun_3$ (\textbf{Invariance to permutation of the endpoints within each measurement}): For each $\MeaInd$, exchange $\TraLoc_\MeaInd$ and $\RecLoc_\MeaInd$ if necessary so that the one closer to the origin appears~first. 

            \item$\InpFun_4$ (\textbf{Invariance to rotation about the vertical axis}): With the $z$ axis pointing upwards,
            rotate every location about this axis so that the horizontal projection of $\RecLoc$ lies on the positive $x$ semi-axis.

            \item$\InpFun_5$ (\textbf{Invariance to mirroring across the $xz$ plane}): Reflect every location across the $xz$ plane if the (soft) majority of $y$-coordinates of the points in $\MeaSet$ is negative.

            \item$\InpFun_{\ComPar, 7}$ (\textbf{Embedding}): Arrange the resulting features into a matrix whose $\MeaInd$-th column collects the transformed coordinates of $\TraLoc_\MeaInd$, $\RecLoc_\MeaInd$, and $\RecLoc$, their magnitudes, their coordinates normalized by their magnitudes, the $z$-coordinate of $\TraLoc$ before $\InpFun_2$, and the measurement value $\Mea_\MeaInd$. Then apply a learnable linear map that embeds each column into $\rfield^{\EmbDim}$, yielding the $\EmbDim\times \MeaNum$ input to $\TraFun$.


        \end{itemize}

    \end{bullets}%

    \blt[Output of the transformer]The only remaining invariance is that to permutations of the measurements. Given the permutation equivariance of $\TraFun$, the entire network is invariant if  $\OutFun$ is properly selected, e.g. as an average of the entries of its input.
    \blt[Learning for variable numbers of measurements]
    \begin{bullets}%
        \blt[average along the embedding dimension]
\blt[causal blocks]However, from a training perspective, it is preferable to use \emph{causal} attention blocks and learn this invariance from data; see~\cite[Sec. IV-C]{viet2025spatial} for details.

        \blt[output dependence on inputs \ra fit all outputs]
        \blt[data augmentation was already going to be performed anyway]
    \end{bullets}%

    The resulting estimator will be referred to as \emph{cross-environment transformer-based estimator} (CRETE).

\end{bullets}%

\section{Numerical Experiments}
\label{sec:experiments}

\begin{bullets}
    \blt[Overview]This section presents simulation experiments that analyze estimation performance in the developed framework. 

    \blt[data]
    \begin{bullets}
        \blt[region]In all experiments, the region $\Reg$ is a volume of $350~\text{m}\times 350~\text{m}\times 20~\text{m}$ and the frequency of interest is $\FreFix=2.4~\text{GHz}$.
        \blt[measurements]Measurements are generated in two ways:
        \begin{bullets}
            \blt[tomographic dataset](i) by ray-tracing in 85 urban environments (17 for testing) with different building layouts. The channel gain is obtained  between all terminals;
            \blt[ray-tracing dataset](ii) via the tomographic model on randomly generated environments that comprise a variable number of buildings whose   horizontal locations are uniformly distributed at random within the limits of $\Reg$. The spatial loss field is defined on a $32\times 32\times 1$ grid and takes the value 1 dB/m inside the buildings and 0 dB/m elsewhere. 50 randomly located terminals are placed in $\Reg$ and the channel gain is computed between all pairs of terminals.
        \end{bullets}%
        \blt[channel parameters]To obtain capacity from channel gain, the bandwidth is set to 20 MHz, the transmit power to 0.3 W, and the noise power to -96 dBm.

    \end{bullets}

    \blt[benchmarks]
        \begin{bullets}
            \blt[tomographic+NN]CRETE is compared with 6 competitors: three tomographic estimators with $\ell_1$~\cite{kanso2009compressed}, total-variation~\cite{wilson2009regularization}, and Tikhonov~\cite{romero2016channelgain} regularization; and a $K$-nearest-neighbor estimator that accounts for both endpoint permutations. Their parameters were tuned for roughly best performance.
            \blt[imai]The estimator from \cite{imai2019radiopredictioncnn} is adapted to cross-environment estimation via~\cite{zintgraf2019cavia}, yielding the explicit estimator IMAI-CAVIA with 80k parameters and a $32\times 32$ context matrix. 
            \blt[Other]A large collection of other explicit estimators were tested, but their performance was not competitive. The best among them, SIREN-CNP, uses the CNP architecture~\cite{garnelo2018cnp} with SIREN activations~\cite{sitzmann2020implicit} and roughly~660k parameters. 
        \end{bullets}%
    \blt[CRETE]CRETE is implemented using a transformer with $L=12$ blocks, 2 attention heads, and embedding dimension $D=128$, yielding roughly 2M parameters. It is trained offline by stochastic minimization of \eqref{eq:implicitcomparopt}, resampling $\TraMeaSet$ and $\ValMeaSet$ at each step; optimizer and schedule details are in the released code. With $\MeaNum$ measurements, per-query complexity is $\mathcal{O}(L\MeaNum^2D+L\MeaNum D^2)$ and attention memory is $\mathcal{O}(L\MeaNum^2)$.

    \blt[experiments]
    \begin{bullets}
        \blt[CG Map estimates]The first experiment compares a true map with its estimate obtained via CRETE. Specifically, the top row of Fig.~\ref{fig:tomographic_cg_est_invariant_transformer}
        shows $\Map(\TraLoc,\RecLoc)$ vs. $\RecLoc$ for four different values of $\TraLoc$. The bottom row is obtained similarly but using the map estimate returned by CRETE. The measurements among the 25 terminal locations were obtained using the aforementioned tomographic model. High similarity between the true and estimated slices can be observed despite the fact that the locations $\TraLoc$ are not among the terminal locations where the measurements were taken.
        \begin{figure}[t]
            \centering
            \includegraphics[width=\columnwidth]{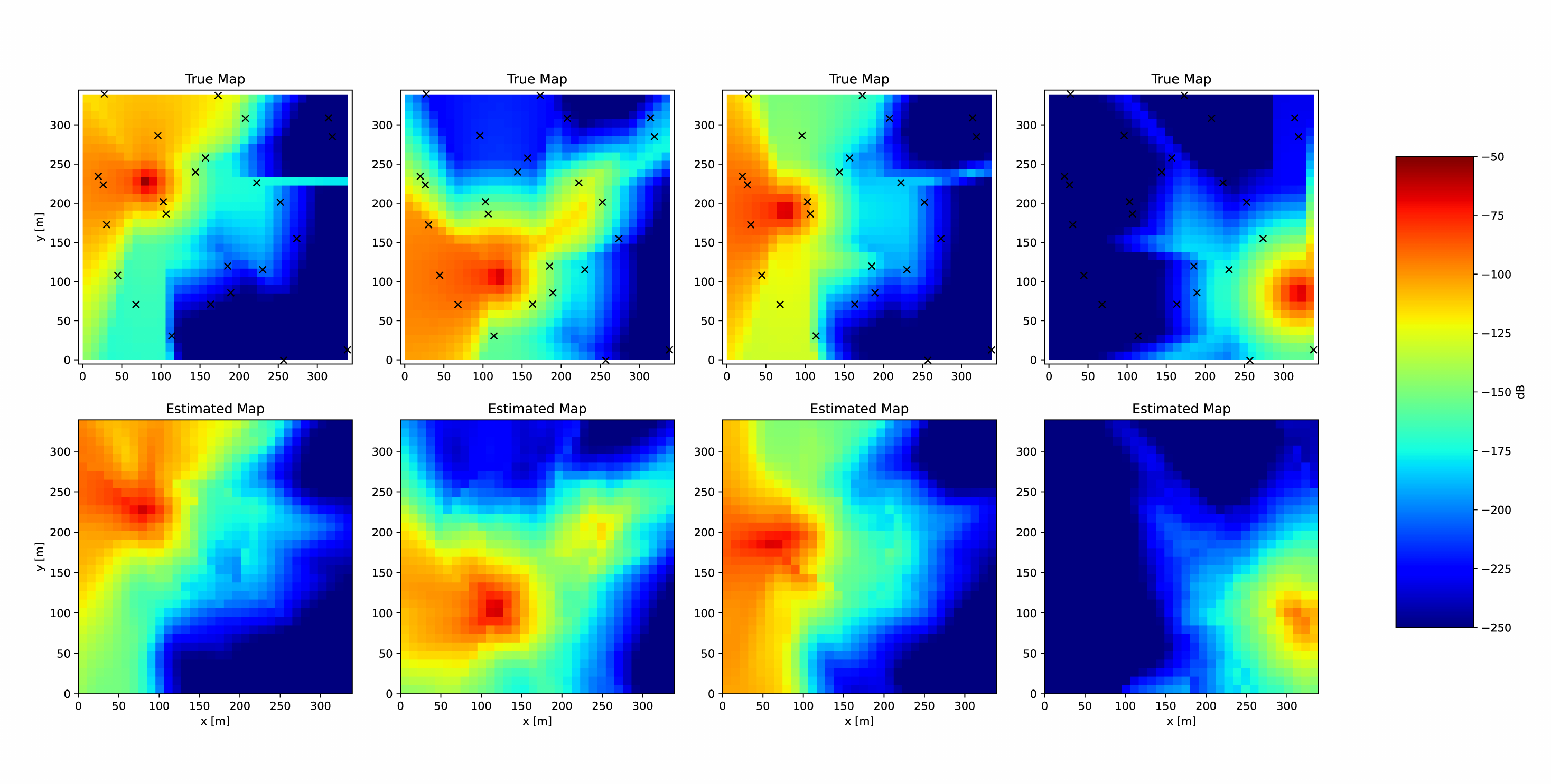}
            \caption{Four slices of the true map (top) and CRETE estimate  (bottom). Each slice (column) corresponds to a different  $\TraLoc$ not present in $\MeaSet$. Terminal locations are shown as black crosses.}
            \label{fig:tomographic_cg_est_invariant_transformer}
        \end{figure}

        \blt[MC: CG Map est. (exp 4152) MAE (dB) vs. Number of measurements]
        \begin{bullets}
            \blt[metric]The second experiment evaluates the \emph{mean absolute error} (MAE), defined as $\expected_{\TraLoc,\RecLoc}|\Map(\TraLoc,\RecLoc)-\MapEst(\TraLoc,\RecLoc)|$, where the expectation is approximated by averaging over 30 pairs $(\TraLoc,\RecLoc)$. 
            \blt[dataset]The MAE is then averaged across scenarios generated from the ray-tracing dataset and shown in Fig.~\ref{fig:rt_compare_cg_est_vs_num_measurements}. 
            \blt[observations]The advantage of the proposed CRETE estimator over the benchmarks is manifest: \emph{achieving, for example, a MAE of 10 dB with CRETE requires 5 times fewer measurements than with the best competitor}. Remarkably, the explicit estimators (IMAI-CAVIA and SIREN-CNP) do not improve significantly as $\MeaNum$ increases, which suggests that they can only exploit simple patterns that can be inferred from few measurements. Increasing model  size did not improve their performance, which suggests that the approximations incurred by explicit estimators (cf. Sec.~\ref{sec:crossenvlearning}) overshadow the benefits of their data-driven nature. 
                        
            \begin{figure}[t]
                \centering
                \includegraphics[width=.8\columnwidth]{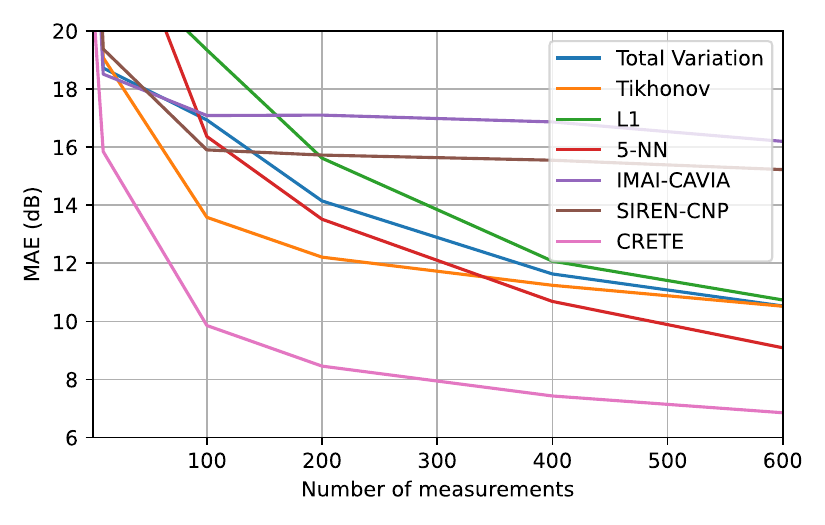}
                \caption{Estimation MAE on the ray-tracing dataset.}
                \label{fig:rt_compare_cg_est_vs_num_measurements}
            \end{figure}
            
        \end{bullets}

        \blt[NMAE of the reconstructed capacity matrix (exp. 4221) vs. the number of
            terminals]
        \begin{bullets}
            \blt[capacity matrix]Establishing ad-hoc networks requires the capacity between all terminal pairs. Estimating the capacity matrix with a subset of the pairwise measurements is one of the applications of CGME; cf. Sec.~\ref{sec:intro}. 
            \blt[dataset+fig]To assess this capability, the normalized MAE (NMAE) of the capacity matrix estimate is plotted in Fig.~\ref{fig:rt_capacity_matrix_vs_num_terminals} vs. the number of terminals in the network when only one half of the pairwise channels are measured. 
            \begin{figure}[t]
                \centering
                \includegraphics[width=.8\columnwidth]{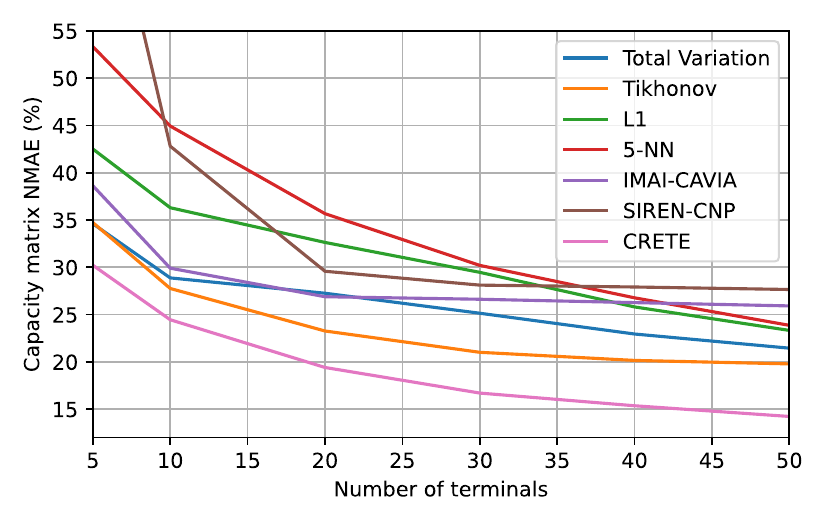}
                \caption{Estimation of the capacity matrix given  half of the measurements between terminal pairs (ray-tracing data).}
                \label{fig:rt_capacity_matrix_vs_num_terminals}
            \end{figure}
            \blt[observations]CRETE exhibits once more a significant advantage over the benchmarks. 
        \end{bullets}

        \blt[MC: Cluster head quality (4321) vs min. capacity for the nearest-neighbor]
        \begin{bullets}
            \blt[metric]In another application of CGME,  the capacity matrix estimate can be used to select a cluster head. For example, one can select the terminal that can reach the largest number of neighbors  with a capacity above a given threshold. 
            \blt[dataset]Fig.~\ref{fig:rt_cluster_head_quality_vs_min_capacity} shows the mean number of neighbors of the cluster head obtained with each estimator normalized by the mean maximum number of neighbors achieved if the capacity matrix were perfectly known. 
            \blt[figure]
            \begin{figure}[t]
                \centering
                \includegraphics[width=.8\columnwidth]{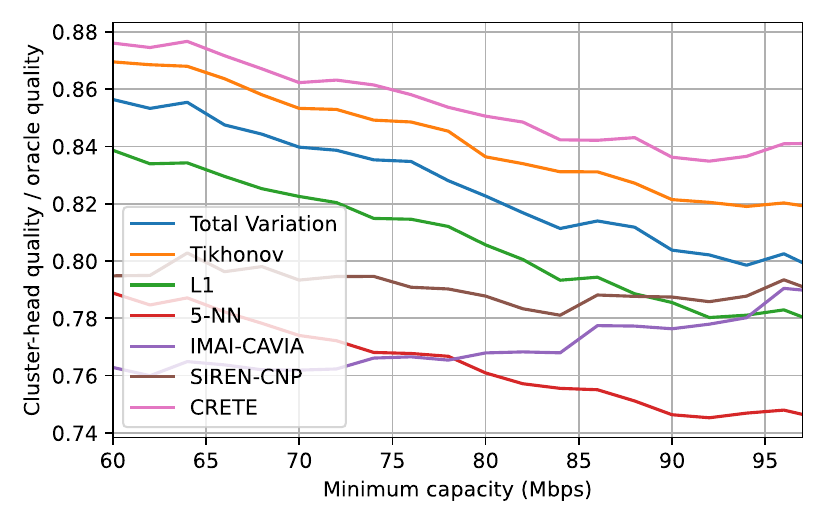}
                \caption{Normalized mean number of neighbors of the cluster head obtained with each estimator (ray-tracing data). }
                \label{fig:rt_cluster_head_quality_vs_min_capacity}
            \end{figure}
            \blt[observations]On average, CRETE achieves a larger number of neighbors than the benchmarks.
        \end{bullets}

        \blt[MC: MAE (4113) vs maximum number of buildings for two MetalearningCgMapEstimators with metanet invariant transformer]
        \begin{bullets}
            \blt[overview]The last experiment quantifies the complexity of the environment. 
            \blt[dataset] To this end, tomographic data is generated with a variable maximum number of buildings. The buildings have the same size but random locations. All true channel-gain maps therefore lie on a manifold of dimension twice the number of buildings.
            \blt[figure]Fig.~\ref{fig:rt_mae_vs_max_num_buildings} shows the MAE of CRETE vs. the maximum number of buildings in the environment.
            \begin{figure}[t]
                \centering
                \includegraphics[width=.8\columnwidth]{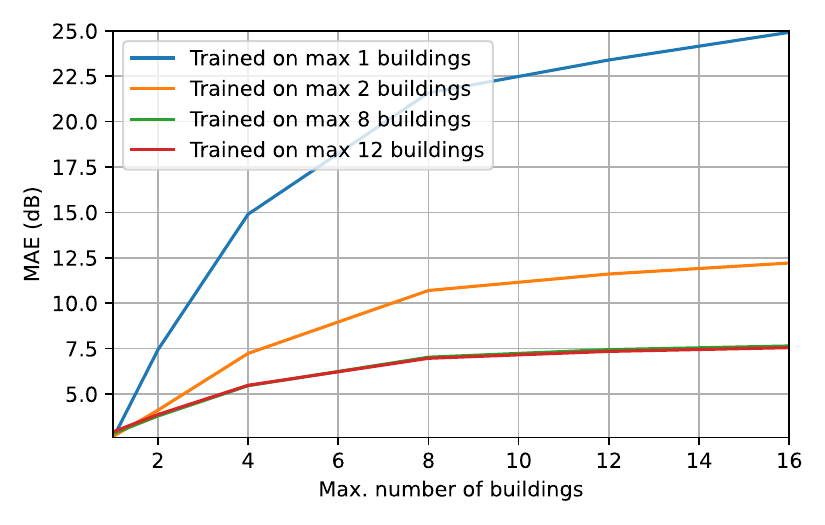}
                \caption{MAE of CRETE vs. the maximum number of buildings.}
                \label{fig:rt_mae_vs_max_num_buildings}
            \end{figure}
            \blt[observations]As anticipated, the MAE increases with the number of buildings, and becomes lower if the training set includes environments with the number of buildings in the test environment. However, diminishing returns are observed: CRETE quickly learns to extrapolate to environments with more buildings than seen during training.
        \end{bullets}

    \end{bullets}
\end{bullets}


\section{Conclusions}
\label{sec:conclusions}
\begin{bullets}%
    \blt[summary]Existing CGME schemes either rely on the (inaccurate) tomographic model or require a prohibitively large number of measurements. 
    \blt[cross-environment]To overcome these limitations, this paper proposed cross-environment learning, where measurements collected in multiple environments are used to implicitly learn the common structure of channel-gain maps in a data-driven fashion.
    \blt[data needs]This approach drastically reduces the number of measurements required to attain a given estimation accuracy. 
    \blt[transformer]To implement this approach, a novel estimator (CRETE) was proposed by relying on the transformer DNN architecture. 
    \blt[future]Future work will validate CRETE with real-world measurements and study generalization across carrier frequencies, terminal heights, and a broader range of propagation environments.
\end{bullets}%

\printmybibliography

\begin{thebibliography}{10}

\bibitem{agrawal2009correlated}
P.~Agrawal and N.~Patwari.
\newblock Correlated link shadow fading in multi-hop wireless networks.
\newblock {\em IEEE Trans. Wireless Commun.}, 8(9):4024--4036, Aug. 2009.

\bibitem{alayafeki2008cartography}
A.~Alaya-Feki, S.~B. Jemaa, B.~Sayrac, P.~Houze, and E.~Moulines.
\newblock Informed spectrum usage in cognitive radio networks: Interference
  cartography.
\newblock In {\em Proc. IEEE Int. Symp. Personal, Indoor Mobile Radio Commun.},
  pages 1--5, Cannes, France, Sep. 2008.

\bibitem{bazerque2010sparsity}
J.-A. Bazerque and G.~B. Giannakis.
\newblock Distributed spectrum sensing for cognitive radio networks by
  exploiting sparsity.
\newblock {\em IEEE Trans. Signal Process.}, 58(3):1847--1862, Mar. 2010.

\bibitem{dallanese2011kriging}
E.~Dall'Anese, S.-J. Kim, and G.~B. Giannakis.
\newblock Channel gain map tracking via distributed kriging.
\newblock {\em IEEE Trans. Veh. Technol.}, 60(3):1205--1211, 2011.

\bibitem{finn2017maml}
C.~Finn, P.~Abbeel, and S.~Levine.
\newblock Model-agnostic meta-learning for fast adaptation of deep networks.
\newblock In {\em International conference on machine learning}, pages
  1126--1135. PMLR, 2017.

\bibitem{garnelo2018cnp}
M.~Garnelo, D.~Rosenbaum, C.~Maddison, T.~Ramalho, M.~Saxton, D.and~Shanahan,
  Y.~W. Teh, D.~Rezende, and SM~A. Eslami.
\newblock Conditional neural processes.
\newblock In {\em International conference on machine learning}, pages
  1704--1713. PMLR, 2018.

\bibitem{ha2024location}
T.~N. Ha, D.~Romero, and R.~López-Valcarce.
\newblock Radio maps for beam alignment in {mmWave} communications with
  location uncertainty.
\newblock In {\em IEEE Veh. Technol. Conf. Workshop}, Singapore, 2024.

\bibitem{imai2019radiopredictioncnn}
T.~Imai, K.~Kitao, and M.~Inomata.
\newblock Radio propagation prediction model using convolutional neural
  networks by deep learning.
\newblock In {\em Proc. IEEE European Conf. Antennas Propag.}, pages 1--5,
  Krakow, Poland, Apr. 2019. IEEE.

\bibitem{iwasaki2020transferbasedpower}
M.~Iwasaki, T.~Nishio, M.~Morikura, and K.~Yamamoto.
\newblock Transfer learning-based received power prediction with ray-tracing
  simulation and small amount of measurement data.
\newblock In {\em IEEE Vehicular Tech. Conf.}, 2020.

\bibitem{kanso2009compressed}
M.~A. Kanso and M.~G. Rabbat.
\newblock Compressed {RF} tomography for wireless sensor networks: Centralized
  and decentralized approaches.
\newblock In {\em Int. Conf. Distributed Comput. Sensor Syst.}, pages 173--186,
  Marina del Rey, CA, 2009. Springer.

\bibitem{kim2011kriged}
S.-J. Kim, E.~Dall'Anese, and G.~B. Giannakis.
\newblock Cooperative spectrum sensing for cognitive radios using {K}riged
  {K}alman filtering.
\newblock {\em IEEE J. Sel. Topics Signal Process.}, 5(1):24--36, Feb. 2011.

\bibitem{kim2013dictionary}
S.-J. Kim and G.~B. Giannakis.
\newblock Cognitive radio spectrum prediction using dictionary learning.
\newblock In {\em Proc. IEEE Global Commun. Conf.}, pages 3206 -- 3211,
  Atlanta, GA, Dec. 2013.

\bibitem{krijestorac2020deeplearning}
E.~Krijestorac, S.~Hanna, and D.~Cabric.
\newblock Spatial signal strength prediction using {3D} maps and deep learning.
\newblock In {\em Proc. IEEE Int Conf. Commun.}, pages 1--6, 2021.

\bibitem{lee2018adaptive}
D.~Lee, D.~Berberidis, and G.~B. Giannakis.
\newblock Adaptive {B}ayesian channel gain cartography.
\newblock In {\em Proc. IEEE Int. Conf. Acoust., Speech, Signal Process.},
  pages 3555--3558, Calgary, Canada, Apr. 2018.

\bibitem{lee2016lowrank}
D.~Lee, S.-J. Kim, and G.~B. Giannakis.
\newblock Channel gain cartography for cognitive radios leveraging low rank and
  sparsity.
\newblock {\em IEEE Trans. Wireless Commun.}, 16(9):5953--5966, Jun. 2017.

\bibitem{lee2024pathloss}
J.-H. Lee and A.~F. Molisch.
\newblock A scalable and generalizable pathloss map prediction.
\newblock {\em IEEE Trans. Wireless Commun.}, 23(11):17793--17806, 2024.

\bibitem{lehmann}
E.~L. Lehmann and G.~Casella.
\newblock {\em Theory of point estimation}.
\newblock Springer, 1998.

\bibitem{levie2019radiounet}
R.~Levie, Ç. Yapar, G.~Kutyniok, and G.~Caire.
\newblock {RadioUNet}: Fast radio map estimation with convolutional neural
  networks.
\newblock {\em IEEE Trans. Wireless Commun.}, 20(6):4001--4015, 2021.

\bibitem{lopezramos2020moe}
L.~M. Lopez-Ramos, Y.~Teganya, B.~Beferull-Lozano, and S.-J. Kim.
\newblock Channel gain cartography via mixture of experts.
\newblock In {\em IEEE Global Commun. Conf.}, pages 1--7, 2020.

\bibitem{pandey2021limited}
A.~Pandey, R.~Sequeira, and S.~Kumar.
\newblock Joint localization and radio map generation using transformer
  networks with limited {RSS} samples.
\newblock In {\em IEEE Int. Conf. Commun. Workshops}, pages 1--6, Jun. 2021.

\bibitem{patwari2008nesh}
N.~Patwari and P.~Agrawal.
\newblock {NeSh}: a joint shadowing model for links in a multi-hop network.
\newblock In {\em Proc. IEEE Int. Conf. Acoust., Speech, Signal Process.},
  pages 2873--2876, Las Vegas, NV, Mar. 2008.

\bibitem{romero2024theoretical}
D.~Romero, T.~N. Ha, R.~Shrestha, and M.~Franceschetti.
\newblock Theoretical analysis of the radio map estimation problem.
\newblock {\em IEEE Trans. Wireless Commun.}, 23(10):13722--13737, Oct. 2024.

\bibitem{romero2022cartography}
D.~Romero and S.-J. Kim.
\newblock Radio map estimation: A data-driven approach to spectrum cartography.
\newblock {\em IEEE Signal Process. Mag.}, 39(6):53--72, 2022.

\bibitem{romero2017spectrummaps}
D.~Romero, S-J. Kim, G.~B. Giannakis, and R.~López-Valcarce.
\newblock Learning power spectrum maps from quantized power measurements.
\newblock {\em IEEE Trans. Signal Process.}, 65(10):2547--2560, May 2017.

\bibitem{romero2016channelgain}
D.~Romero, D.~Lee, and G.~B. Giannakis.
\newblock Blind radio tomography.
\newblock {\em IEEE Trans. Signal Process.}, 66(8):2055--2069, Apr. 2018.

\bibitem{romero2022aerial}
D.~Romero, P.~Q. Viet, and G.~Leus.
\newblock Aerial base station placement leveraging radio tomographic maps.
\newblock In {\em IEEE Int. Conf. Acoustics Speech Signal Process.}, pages
  5358--5362, Singapore, 2022. IEEE.

\bibitem{schaufele2019tensor}
D.~Sch\"{a}ufele, R.~L.~G. Cavalcante, and S.~Mtanczak.
\newblock Tensor completion for radio map reconstruction using low rank and
  smoothness.
\newblock In {\em Proc. IEEE Workshop Signal Process. Adv. Wireless Commun.},
  Cannes, France, Jul. 2019.

\bibitem{scholkopf2001}
B.~Schölkopf and A.~J. Smola.
\newblock {\em Learning with Kernels: Support Vector Machines, Regularization,
  Optimization, and Beyond}.
\newblock MIT Press, 2002.

\bibitem{shrestha2022surveying}
R.~Shrestha, D.~Romero, and S.~P. Chepuri.
\newblock Spectrum surveying: Active radio map estimation with autonomous
  {UAVs}.
\newblock {\em IEEE Trans. Wireless Commun.}, 22(1):627--641, 2022.

\bibitem{sitzmann2020implicit}
V.~Sitzmann, J.~Martel, A.~Bergman, D.~Lindell, and G.~Wetzstein.
\newblock Implicit neural representations with periodic activation functions.
\newblock {\em Advances in Neural Inf. Process. Systems}, 33:7462--7473, 2020.

\bibitem{sun2025scatterer}
H.~Sun, L.~Zhu, and R.~Zhang.
\newblock Channel gain map estimation for wireless networks based on scatterer
  model.
\newblock {\em IEEE Trans. Wireless Commun.}, 24(8):7012--7028, 2025.

\bibitem{teganya2020rme}
Y.~Teganya and D.~Romero.
\newblock Deep completion autoencoders for radio map estimation.
\newblock {\em IEEE Trans. Wireless Commun.}, 21(3):1710--1724, 2021.

\bibitem{teganya2019locationfree}
Y.~Teganya, D.~Romero, L.~M. Lopez-Ramos, and B.~Beferull-Lozano.
\newblock Location-free spectrum cartography.
\newblock {\em IEEE Trans. Signal Process.}, 67(15):4013--4026, Aug. 2019.

\bibitem{vaswani2017attention}
A.~Vaswani, N.~Shazeer, N.~Parmar, J.~Uszkoreit, L.~Jones, A.~N. Gomez,
  L.~Kaiser, and I.~Polosukhin.
\newblock Attention is all you need.
\newblock {\em Advances neural inf. process. systems}, 30, 2017.

\bibitem{viet2025spatial}
P.~Q. Viet and D.~Romero.
\newblock Spatial transformers for radio map estimation.
\newblock In {\em IEEE Int. Conf. Commun.}, Montreal, Canada, 2025.

\bibitem{wang2023bert}
Z.~Wang, Q.~Kong, B.~Wei, L.~Zhang, and A.~Tian.
\newblock Radio map construction based on {BERT} for fingerprint-based indoor
  positioning system.
\newblock {\em Eurasip J. Wirel. Commun. Netw.}, 2023(1):39, 2023.

\bibitem{wilson2009regularization}
J.~Wilson, N.~Patwari, and O.~G. Vasquez.
\newblock Regularization methods for radio tomographic imaging.
\newblock In {\em Virginia Tech Symp. Wireless Personal Commun.}, Blacksburg,
  VA, Jun. 2009.

\bibitem{zintgraf2019cavia}
L.~Zintgraf, K.~Shiarli, V.~Kurin, K.~Hofmann, and S.~Whiteson.
\newblock Fast context adaptation via meta-learning.
\newblock In {\em Int. Conf. Machine Learning}, pages 7693--7702. PMLR, 2019.

\end{thebibliography}
\end{document}